\documentstyle[12pt,fleqn]{article}

\title{\Large Emission of gamma rays by\\
X-ray electron-nuclear double transitions}

\author{Silviu Olariu, Agata Olariu and Valeriu Zoran\\
Institute of Physics and Nuclear Engineering, Heavy-Ion Physics Department\\
76900 Magurele, P.O. Box MG-6, Bucharest, Romania\\}

\begin{document}
\date{}
\maketitle

\abstract
The X-ray electron-nuclear double transitions (XENDT) are processes in which 
a transition effected by an inner atomic electron takes place simultaneously
with a nuclear electromagnetic transition. We give expressions for the
cross sections of electric and magnetic XENDTs of various multipole orders.
We calculate the rate of deexcitation of isomeric nuclei induced by XENDTs 
for the case when the holes in the atomic shells are produced by incident 
ionizing electrons and find that the induced nuclear deexcitation rate 
becomes comparable to the natural decay rate for ionizing fluxes of the
order of $10^{14}$ W cm$^{-2}$. We show that for E1 and M1 nuclear
processes for 
which there is a matching between the electron and the nuclear transition 
energies, the XENDTs can be used to produce pulses of M\"{o}ssbauer radiation, 
with yields of the order of $10^4$ Bq mA$^{-1}$. \\
\endabstract

\setlength{\parindent}{0cm}
\newpage

The electron-nuclear double transitions are a class of processes in 
which a transition effected by an atomic electron takes place simultaneously 
with a nuclear electromagnetic transition. There are several types of 
electron-nuclear double transitions,
depending on the range of energies for the electron and nuclear transition
energies. Thus,
the application of a radiofrequency magnetic field to a sample generates 
additional lines in the M\"{o}ssbauer spectra \cite{1}-\cite{7}.
In the case of the electron-nuclear double resonance \cite{8}-\cite{10},
a microwave magnetic field 
interacts with the Zeeman sublevels of electrons from incomplete atomic shells 
while a radiofrequency magnetic field interacts resonantly with the magnetic
sublevels of the ground nuclear state. Another
possibility \cite{11} is that the electron transition takes place between the 
Zeeman
substates of incomplete electron shells while the nucleus makes a gamma-ray
transition. These electron-nuclear double transitions produce changes
in the M\"{o}ssbauer spectra, and may have applications for the research on the
amplification of gamma rays without inversion of nuclear population, and for 
the identification of the position of lines in complex M\"{o}ssbauer spectra.
In the case of the electron-nuclear double transitions at optical frequencies, 
the optical
pumping of atomic hyperfine transitions by a laser induces an anisotropy in the
angular distribution of gamma rays emitted by the nucleus \cite{12}.\\

In the case of X-ray electron-nuclear double transitions (XENDTs), the 
electrons make transitions to fill the holes created in the atomic shells by a
beam of incident electrons or by the electrons of a dense hot plasma. In the 
process of nuclear excitation by electron transition  
first described by Morita \cite{13}-\cite{21},
an electron from a higher electron shell makes a transition to fill a hole in
an inner electron shell while at the same time the nucleus makes a transition
from the ground nuclear state to an excited nuclear state. The nuclear 
excitation by electron transition competes with the conventional atomic 
deexcitation via X-ray emission or Auger emission. The branching ratio for the
nuclear excitation by electron transition depends on the energy of interaction
between the inner-shell electron and the protons of the nucleus, and since the
inner electrons are rather close to the nucleus, this energy of interaction is
relatively large. The branching ratio for the nuclear excitation by electron
transition is inversely proportional to the square of the difference between
the electron transition energy and the nuclear transition energy. Therefore,
the nuclear excitation by electron transition becomes important when there is
a matching between the transition energy from a populated nuclear state, like
the ground state or an isomeric state, and an X-ray transition energy. Typical
transition energies are of the order of a few tens of keV's, and the detuning 
between the electron and nuclear transition energies may be of about 1 keV. 
The 
nucleus can also make a transition from an excited state to a lower state while
an electron is raised from an inner shell to a hole in a higher shell. This
process may be called nuclear deexcitation by electron transition.
The process of nuclear excitation by electron transition has been observed
in $^{189}$Os \cite{14}, \cite{15}, \cite{19}, $^{235}$U \cite{16}, 
$^{237}$Np \cite{17} and $^{197}$Au \cite{20}. Although the cross sections for
the nuclear excitation by electron transitions are rather small, these
processes are interesting because they represent a way to control a nuclear 
transition by actions exerted at the atomic level.\\

In this work we give expressions for the cross sections of electric or
magnetic XENDTs, and study the application of XENDTs to the problem of induced
gamma emission. If the nucleus is initially in a long-lived isomeric state
$|i\rangle$, the XENDTs open a new deexcitation channel for the isomeric 
state, 
in addition to the regular gamma-ray emission and internal conversion. One 
possibility is the direct deexcitation of the isomeric nucleus by
a transition from the isomeric state $|i\rangle$ to a lower state $|l\rangle$,
while an electron is raised to fill 
a hole in a higher electron shell. Another possibility is the 
two-step deexcitation of the isomeric nucleus by a
transition from the isomeric state $|i\rangle$ to a higher nuclear state 
$|h\rangle$ while
an electron from a higher shell makes a transition to a hole in an inner
electron shell .
The nuclear intermediate state $|h\rangle$ then decays by the 
emission of a gamma-ray photon to the lower nuclear state $|l\rangle$. These 
alternative nuclear paths for the deexcitation of an isomeric state by 
electron transition are shown in Fig. 1.  The XENDT rate is proportional to the
electron flux which produces the holes in the atomic shells, and for 
sufficiently large ionizing electron fluxes the deecitation rate induced by
XENDT becomes comparable to the natural decay rate of the isomer. Another
application of XENDTs is the possibility of producing pulsed M\"{o}ssbauer
gamma radiation for isotopic elements where there is a matching between an
electron transition energy and a resonant nuclear transition energy.\\

We shall calculate the probability of an XENDT for electric or magnetic
interactions of arbitrary multipole orders, and shall estimate the cross
section for the production of an XENDT for the case when the holes in the
atomic shells are produced by incident ionizing electrons.  
We shall estimate the ionizing electron fluxes for which the XENDT rate 
becomes
equal to the natural decay rate of the isomeric state, both for a direct 
deexcitation $|i\rangle\rightarrow
|l\rangle$ and for a two-step deexcitation $|i\rangle\rightarrow 
|h\rangle\rightarrow |l\rangle$. Then we shall  list the
isotopic species for which it is possible to generate resonant M\"{o}ssbauer
radiation by XENDT, and shall estimate the corresonding yields.\\

The initial state of the XENDT process is prepared when an 
electron hole is created in a certain atomic subshell, for example by electron
impact. We designate this initial 
atomic state by $|a_1\rangle$ and its energy by $E_1^{(a)}>0$, as shown in Fig.
2. Moreover, we designate the initial nuclear state by $|n_1\rangle$ and the 
energy of the initial nuclear state by $E_1^{(n)}$. The initial state of the
XENDT process is thus $|a_1n_1\rangle$. In the course of the 
XENDT, an electron from another atomic subshell fills the initial hole, 
thereby creating a new
hole in a different subshell and thus leaving the atom in the final state 
$|a_2\rangle$
of energy $E_2^{(a)}>0$. The electron transition $|a_1\rangle\rightarrow 
|a_2\rangle$ takes place simultaneously with a
nuclear transition $|n_1\rangle\rightarrow |n_2\rangle$ from the initial
nuclear state $|n_1\rangle$ to the final nuclear state $|n_2\rangle$ of energy
$E_2^{(n)}$. The final state of the XENDT process is thus $|a_2n_2\rangle$.
If $E_1^{(n)}>E_2^{(n)}$, we have a nuclear deexcitation by 
electron transition, as shown in Fig. 2(a), and if $E_1^{(n)}<E_2^{(n)}$ 
we have a nuclear excitation by electron transition, as shown
in Fig. 2(b).\\

The probability of the XENDT depends on the detuning $\Delta$ between the
total initial energy $E_1^{(n)}+E_1^{(a)}$ and the total final energy
$E_2^{(n)}+E_2^{(a)}$,
\begin{equation}
\Delta = E_1^{(n)}+E_1^{(a)}-(E_2^{(n)}+E_2^{(a)}) .
\end{equation}
The widths $\Gamma_1, \Gamma_2$ of the atomic states $|a_1\rangle, |a_2\rangle$
are determined by the X-ray emission and the Auger and Coster-Kronig emission
of electrons.
We shall represent these processes by imaginary terms
$-i\Gamma_1/2, -i\Gamma_2/2$ in the Hamiltonian matrix of the electron-nuclear
system. If $c_1(t), c_2(t)$ are the amplitudes to find the electron-nuclear
system in the states $|a_1n_1\rangle$ and respectively $|a_2n_2\rangle$, the 
Hamiltonian equations for this system are \cite{21b}
\begin{equation}
i\hbar\frac{dc_1}{dt}=-\frac{i\Gamma_1}{2} c_1 + V_{12}\: e^{i\Delta t/\hbar}
\:c_2 ,
\end{equation}
\begin{equation}
i\hbar \frac{dc_2}{dt}=V_{12}^*\:e^{-i\Delta t/\hbar}c_1-\frac{i\Gamma_2}
{2}c_2 ,
\end{equation}
where $V_{12}=\langle a_1n_1|V|a_2n_2\rangle$ is the matrix element for the 
interaction
between the electrons and the nucleus, and $V_{12}^*$ is the complex conjugate
of $V_{12}$. 
The solution of Eqs. (2), (3) with the initial conditions
$c_1=1, c_2=0$ at
$t$=0 is
\begin{equation}
c_1(t)=e^{i\Delta t/\hbar-(\Gamma_1+\Gamma_2)t/4\hbar}\:\left(\cos\Omega t -
\frac{i\Delta +(\Gamma_1-\Gamma_2)/2}{2\hbar \Omega}\sin\Omega t\right)
\end{equation}
\begin{equation}
c_2(t)=-\frac{iV_{12}^*}{\hbar \Omega}\:e^{-i\Delta t/\hbar-
(\Gamma_1+\Gamma_2)t/4\hbar}\:\sin\Omega t ,
\end{equation}
where
\begin{equation}
\Omega^2=\frac{1}{4\hbar^2}\left(\Delta-i\frac{\Gamma_1-\Gamma_2}{2}\right)^2+
\frac{\vert V_{12}\vert^2}{\hbar^2}.
\end{equation}
The fact that $c_1(t)\rightarrow 0, c_2(t)\rightarrow 0$ as
 $t\rightarrow \infty$ is due to the presence of the imaginary diagonal terms
in the Hamiltonian equations (2), (3).\\

The probability $P$ to have the nucleus in the state $n_2$ at the end of the 
process can be obtained by multiplying the probability $|c_2(t)|^2$ of finding
the electron-nuclear system in the state $|a_2n_2\rangle$ by the probability 
$\Gamma_2 dt$ of an electron transition from the state $|a_2n_2\rangle$ 
to other states $|a_2^\prime n_2\rangle$ ,
\begin{equation}
P=\int_0^\infty \vert c_2(t)\vert^2 \Gamma_2 dt .
\end{equation}
Usually we have $|\Delta|\gg \Gamma_1,\Gamma_2, |V_{12}|$, so that
$\Omega^2\approx \Delta^2/4\hbar^2,$ and then
\begin{equation}
P=\frac{\Gamma_1+\Gamma_2}{\Gamma_1}\frac{|V_{12}|^2}{\Delta^2} .
\end{equation}
This is in agreement with the expression given in \cite{15}. 
The $1/\Delta^2$ dependence is characteristic to non-resonant Rabi oscillations
in two-state systems. \cite{21c}
The phase of the state $|a_1n_1\rangle$ is the sum of the phase
of the atomic state $|a_1\rangle$ and of the phase of the 
nuclear state $|n_1\rangle$, and
similarly the phase of the state $|a_2n_2\rangle$ is the sum of the phase
of the atomic state $|a_2\rangle$ and of the phase of the
nuclear state $|n_2\rangle$.
The probability $P$, Eq. (8), is independent of these phases.\\

We shall assume  that in the initial atomic state $|a_1\rangle$ the hole is 
in the
$nl_J$ subshell of principal quantum number $n$, orbital angular momentum
$l$ and total angular momentum $J$. An atomic electron can make a transition
to fill this hole, thereby creating the final state $|a_2\rangle$ having a 
hole in the
subshell  $n^\prime l^\prime_{J^\prime}$ . If the angular momentum and
parity of the nuclear states $|n_1\rangle,|n_2\rangle$ are respectively 
$I_1^{\pi_1},
I_2^{\pi_2}$,  then the angular momentum $F_1$ of the state $|nl_J,n_1\rangle$ 
is such that $|J-I_1|<F_1<J+I_1$ and the parity of the state $|nl_J,n_1\rangle$
is $\Pi_1=\pi_1 (-1)^l$.  The angular momentum $F_2$ of the state 
$|n^\prime l^\prime_{J^\prime},n_2\rangle$ 
is such that $|J^\prime-I_2|<F_2<J^\prime+I_2$ 
and the parity of the state $|n^\prime l^\prime_{J^\prime},n_2\rangle$
is $\Pi_2=\pi_2 (-1)^{l^\prime}$.\\

In Eq. (8) we have assumed that {\it one} electron interacts with the nucleus 
to produce the XENDT. There are however $2J^\prime+1$ electrons which can make
a transition from the $n^\prime l^\prime_{J^\prime}$ subshell to fill
the hole in the $nl_J$ subshell, so that the total transition probability
is
\begin{equation}
P_{tot}=(2J^\prime+1)P .
\end{equation}

We shall estimate the matrix element $V_{12}$ appearing in Eq. (8) as the
energy of interaction between two electric multipoles of the same order,
or between two magnetic multipoles of the same order. In the case of the
interaction between electric multipoles of order $L$, the energy of 
interaction \cite{22}, \cite{23} can be estimated as
\begin{equation}
V_{12}^{({\rm E}L)}=\frac{3 f_E e^2 r_A^L \langle n^\prime
l^\prime_{J^\prime}\vert r^{-L-1}\vert nl_J\rangle}{(4\pi)^{1/2}\epsilon_0 
(2L+1)(L+3)} ,
\end{equation}
where $r_A$ is the radius of a nucleus of mass number $A$, $r_A=r_0A^{1/3},
r_0=1.2\cdot 10^{-15}$ m,  and the dimensionless factor $f_E$ is of the order
of unity. In the independent-particle nuclear model used in this work the
energy of interaction $V_{12}^{(EL)}$ is not proportional to the proton number
$Z$.\\

The energy of interaction between two magnetic multipoles of order $L$ can
be estimated as 
\begin{equation}
V_{12}^{({\rm M}L)}=f_M\left(\frac{\hbar}{m_pcr_A}\right)
\frac{3 e^2 r_A^L}{(4\pi)^{1/2}\epsilon_0 (2L+1)(L+3)} 
\langle n^\prime l^\prime_{J^\prime}\vert 
\left(m_e^2 c^2 r^2/\hbar^2+1\right)^{-1/2} r^{-L-1}\vert nl_J\rangle  ,
\end{equation}
where $m_e$ is the electron mass, $m_p$ the proton mass,
and the dimensionless factor $f_M$ has values of a few units. The factor
$\hbar/m_pcr_A$ is of the order of $v_p/c$, where $v_p$ is the proton velocity,
and gives approximately the ratio between the magnetic and the electric
multipole moments,\cite{23} and the factor 
$\left(m_e^2 c^2 r^2/\hbar^2+1\right)^{-1/2}$
is of the order of $v_e(r)/c$, where $v_e(r)$ is the electron velocity
at a distance $r$ from the origin. For $r\gg \hbar/m_ecr$ the factor
$(m_e^2c^2r^2/\hbar^2+1)^{-1/2}$ is approximately equal to $\hbar/m_ecr$, while
for $r\rightarrow 0$ it converges to 1.  \\

The total transition probability in the electric case is then
\begin{equation}
P_{tot}^{({\rm E}L)}=\frac{\Gamma_1+\Gamma_2}{\Gamma_1}\frac{(2J^\prime+1)
\vert V_{12}^{({\rm E}L)}\vert^2}
{\Delta^2} ,
\end{equation}
and the total transition probability in the magnetic case is
\begin{equation}
P_{tot}^{({\rm M}L)}=\frac{\Gamma_1+\Gamma_2}{\Gamma_1}\frac{(2J^\prime+1)
\vert V_{12}^{({\rm M}L)}\vert^2}{\Delta^2} .
\end{equation}
In this work we shall assume that $f_E$=1, $f_M^2=10$, as in the
Weisskopf estimate of gamma-ray transition rates.\cite{23}
As in the case of the
Weisskopf estimates, the probabilities in Eqs. (12), (13) may differ from
the real transition probabilities by about two orders of magnitude.\\

The cross section of an XENDT is then
\begin{equation}
\sigma=\sigma_{hole} P_{tot} ,
\end{equation}
where $\sigma_{hole}$ is the cross section for the production of the initial
hole in the atomic subshell $nl_J$. We shall estimate $\sigma_{hole}$ according
to Gryzi\'{n}ski \cite{24} as
\begin{equation}
\sigma_{hole}=\frac{\pi e^4 (2J+1)}{16\pi^2\epsilon_0^2 E_{nl_J}^2}
g_i(E_{el}/E_{nl_J}) ,
\end{equation}
where $E_{el}$ is the energy of the ionizing electron, and where the
expression of the function $g_i$ is given in ref. \cite{24}. We shall assume
that $E_{el}=1.6 E_{nl_J}$, and we have $g_i(1.6)=0.109$.\\

In the case of a direct nuclear transition from state $|i\rangle$ to state
$|l\rangle$ the XENDT rate is
\begin{equation}
R_I=\sigma N_{el} ,
\end{equation}
where $N_{el}$ is the number of ionizing electrons per unit surface
and unit time. The relative deexcitation rate for an isomeric state of 
half-life $t_i$ is then
\begin{equation}
T_I=\sigma N_{el}t_i/\ln 2 .
\end{equation}
In the case of a two-step nuclear transition $|i\rangle\rightarrow |h\rangle
\rightarrow |l\rangle$, the induced deexcitation rate of the state 
$|i\rangle$ is
\begin{equation}
R_{II}=\sigma N_{el}B ,
\end{equation}
where $B$ is the branching ratio for the gamma-ray transition $|h\rangle
\rightarrow |l\rangle$. 
We have determined $B$ from the Weisskopf estimate of the radiative widths
$\Gamma_{hi}, \Gamma_{hl}$ and from the internal conversion coefficients
$\alpha_{hi}, \alpha_{hl}$ of the transitions $|h\rangle\rightarrow |i\rangle, 
|h\rangle \rightarrow |l\rangle$ as
\begin{equation}
B=\frac{(1+\alpha_{hl})\Gamma_{hl}}{(1+\alpha_{hi})\Gamma_{hi}+
(1+\alpha_{hl})\Gamma_{hl}} .
\end{equation}
The relative deexcitation rate for an isomeric half-life $t_i$ is then
\begin{equation}
T_{II}=\sigma N_{el}Bt_i/\ln 2.
\end{equation}

The induced gamma emission becomes significant when the
induced and natural decay rates are equal, so that $T_{I}=1$ or $T_{II}=1$ .
In the case of a direct deexcitation process the energy flux $\Phi_I$ 
of the ionizing electrons for which  
$T_{I}=1$  is
\begin{equation}
\Phi_I=\frac{\ln 2 E_{el}}{\sigma t_i} .
\end{equation}
In the case of a two-step deexcitation process the energy flux $\Phi_{II}$ 
of the ionizing electrons for which  
$T_{II}=1$  is
\begin{equation}
\Phi_{II}=\frac{\ln 2 E_{el}}{\sigma t_i B} .
\end{equation}

In Table I we have given the results of calculations on the direct
deexcitation of isomeric nuclei induced by XENDTs, for several isomeric 
nuclides
having a half-life $t_i>10$ minutes, and for which there is a downward
transition $|i\rangle\rightarrow |l\rangle$ of energy $E_{il}<100 $ keV. We 
have used the total atomic level widths as given by Keski-Rahkonen and Krause,
\cite{25} which include the radiative width, the Auger width and the 
Coster-Kronig width.
In order to evaluate the matrix elements in Eqs. (10) and (11) we have
represented the atomic states 
$|nl_J\rangle, |n^\prime l^\prime_{J^\prime}\rangle$
by screened hydrogenic wave functions. The values of the screening constants
have been determined according to the rules of Slater. \cite{25b}
In Tables I, II and III  we have used the values of the gamma-ray 
transition energies, energy levels and half-lifes adopted by the IAEA
Nuclear Data Information System, and for the atomic energy levels we have
used the values listed by Lederer et al. \cite{25c}
We see from Table I that for direct induced emission the lowest 
values of the ionizing electron energy  flux $\Phi_I$ are of the order of 
$10^{15}$ W cm$^{-2}$, for $^{174m}$Lu and $^{99m}$Tc.\\

In Table II we have given the results of calculations on the two-step 
deexcitation of isomeric nuclei induced by XENDTs, for several isomeric 
nuclides having a half-life $t_i>10$ minutes, and for which there is an upward
transition $|h\rangle\rightarrow |i\rangle$ of energy $E_{hi}<100 $keV. 
We have used the
theoretical internal conversion coefficients of Band et al. \cite{26} and
of R\"{o}ssel et al. \cite{27} The values of the branching ratio $B$ are
calculated according to  Eq. (19).
We see from Table II that for the two-step 
induced emission the lowest values of
the ionizing electron energy  flux $\Phi_{II}$ are of the order of 
$10^{14}$ W cm$^{-2}$, for  $^{174m}$Lu and $^{188m}$Re.\\

In the case of $^{99m}$Tc the energy of the isomeric transition is $E_{il}$=2.1
keV, and the  $^{99}$Tc nucleus makes a transition from the $|l\rangle$ level 
to the ground state with the emission of an 140.5 keV photon.  In the case of 
$^{188m}$Re
the energy of the upward transition is $E_{hi}$=10.7 keV, then the $^{188}$Re
nucleus emits 26.7 keV, 63 keV, 156 keV photons. In the case of $^{174m}$Lu
the energy of the upward transition is $E_{hi}$=29.1 keV, then the $^{174}$Lu
nucleus emits 44.7 keV, 67.1 keV, 88.2 keV photons. Thus, the emission of
gamma rays induced by XENDT may be regarded in these cases 
as an upconversion of the incident
electron energy. The efficiency of the upconversion process is however 
very low.\\

A preliminary step in the direction of the gamma-ray emission from nuclear 
isomers induced by XENDTs would be the generation of resonant M\"{o}ssbauer
gamma radiation by XENDTs. In this case, a transition of a nucleus to an 
excited state is induced by an electron transition to a hole in an inner
atomic shell, the hole being produced by incident ionizing electrons. The 
nucleus thus excited then emits a gamma-ray photon whose energy is narrowly
centered on the nuclear transition energy, and can be used for regular
M\"{o}ssbauer experiments. This type of M\"{o}ssbauer source would be active
only as long as it is excited by the incident electrons.\\

If ionizing electrons of suitable energy are incident on a thin layer of
surface $S$ and thickness $d_e$ containing the XENDT nuclei, the number
$N_\gamma$ of resonant gamma-ray photons emitted per second by the nuclei
in the layer is
\begin{equation}
N_\gamma=\frac{INd_e}{(1+\alpha)e}\sigma ,
\end{equation}
where $I$ is the incident electron current, $e>0$ is the electron charge,
$N$ the concentration of XENDT nuclei in the foil, and $\alpha$ the internal
conversion coefficient for the transition under study. The number of
M\"{o}ssbauer gamma-ray photons generated per second via XENDTs per unit
of incident electron current is then $N_\gamma/I$. We have estimated the
thickness $d_e$ as
\begin{equation}
d_e=a\frac{A}{Z\rho}E_{nl_J}^2 ,
\end{equation}
where $a=5.63\cdot10^{-3}$ when $d_e$ is in $\mu$m, the density $\rho$ of
the target in g cm$^{-3}$ and $E_{nl_J}$ in keV, $A$ and $Z$ being the
mass number and the proton number for the nuclei in the target.\\

In Table III we have listed the stable nuclides and the nuclides having a
half-life greater than 100 days for which the detuning is
$|\Delta|<10$ keV. We have also required that the spectral gamma-ray intensity
at the center of the M\"{o}ssbauer line produced by XENDT should be greater 
than the spectral
intensity of the bremsstrahlung generated by the ionizing electrons, and should
also be greater than the spectral intensity due to non-resonant X-ray emission.
We have used the X-ray emission rates calculated by Scofield. \cite{28}
The afore-mentioned conditions are fulfilled by E1 and M1
transitions when there is a matching between the X-ray transition energy
and the nuclear transition energy. We have evaluated the induced M\"{o}ssbauer
activity per unit of incident electron current $N_\gamma/I$ for a target
concentration $N=10^{22}$ nuclides cm$^{-3}$. The nuclides in Table III which
have a significant recoilless fraction at room temperature are $^{119}$Sn,
$^{161}$Dy, $^{189}$Os, $^{193}$Ir. 
For the E1 and M1 transitions listed in Table III, the energy of interaction
$V_{12}$ is of the order of 1 eV, the probability $P_{tot}$ of the XENDT
process is in the $10^{-7}\cdot 10^{-4}$ range, the cross section of the XENDT
process with respect to the incident ionizing electrons is in the range
$10^{-30}\cdots 10^{-28}$ cm$^2$, and the M\"{o}ssbauer
activity per unit of incident electron current $N_\gamma/I$ is of the order
of $10^{4}$ Bq/mA. This would be a rather weak continuous gamma-ray source, 
but the interesting property of such a M\"{o}ssbauer source
is that it can produce {\it pulses} of M\"{o}ssbauer radiation.\\

We have assumed so far that the holes in the atomic shells are produced by
electrons. The generation of M\"{o}ssbauer radiation by XENDTs can be
also be studied with incident protons having an energy of several MeV.
The smaller cross sections for the generation of the holes in the atomic
shells are compensated by the larger proton range in the target.\\

In this paper we have investigated the possibility of  inducing nuclear 
deexcitation rates
comparable to the natural emission rates from nuclear isomeric states
by using the relatively large energy of interaction extant between the
nuclear protons and the inner atomic electrons. 
The calculations of the present
work show that significant induced-emission rates could be obtained for
ionizing energy fluxes  of the order of $10^{14}$ W cm$^{-2}$. 
An application of XENDTs which  requires
lower electron fluxes is the generation for certain elements
of pulses of M\"{o}ssbauer gamma radiation.\\

ACKNOWLEDGMENT\\

This work has been supported by a research grant from the Romanian Ministry of
Research and Technology.\\

\newpage
FIGURE CAPTIONS\\

Fig. 1. Nuclear paths for the deexcitation of an isomeric state by electron
transition. The two possibilities are the direct deexcitation from the 
isomeric state $|i\rangle$ to a 
lower state $|l\rangle$ via a nuclear deexcitation by electron transition,
and the two-step deexcitation, when the
nucleus goes from the isomeric state $|i\rangle$ to a higher state nuclear 
state $|h\rangle$ via a nuclear excitation by electron transition,  
then it decays to the lower state $|l\rangle$ by regular gamma-ray emission.\\

Fig. 2. The two cases of XENDT. (a) Nuclear deexcitation
by electron transition (NDET), when an electron from an inner shell makes
a transition to fill a hole in a higher electron shell. (b) Nuclear excitation
by electron
transition (NEET) when an electron from a higher electron shell makes a
transition to fill a hole in an inner shell.  The subscripts 1, 2 indicate
the initial state and respectively the final state. The higher electron
shells are represented by the upper lines in the atomic level schemes.
The arrows in the atomic schemes suggest the path of the electron which
fills the hole in the initial state.\\

\newpage
Table I. Direct deexcitation of isomeric nuclei induced by XENDT.\\

\begin{tabular}{|c|c|c|c|c|c|c|c|c|c|c|c|}
\hline
nuclide&$E_{il}$,&E$L$/&$t_i$, s&$nl_J$&$n^\prime
l^\prime_{J^\prime}$&$V_{12}$, &$\Delta$, &$P_{tot}$&$\sigma$, 
&$E_{el}$, &$\Phi_I,$\\
&keV&M$L$&&&&eV&keV&&cm$^2$&keV&W cm$^{-2}$\\
\hline
$^{85}$Sr &6.96 &E3&4.06E3 &$3d_{5/2}$&$2p_{3/2}$&3.27E-9 & 5.15 &3.12E-23&
7.50E-41&0.213&7.76E19\\
$^{86}$Y  &10.2 &E3&2.88E3 &$3d_{5/2}$&$2p_{3/2}$&4.04E-9 & 8.23 &2.06E-23&
3.51E-41&0.253&2.78E20\\
$^{99}$Tc &2.17 &E3&2.16E4 &$3d_{5/2}$&$2p_{3/2}$&9.47E-9 &-0.251&6.30E-20&
4.19E-38&0.405&4.96E16\\
$^{162}$Ho&10   &E3&4.02E3 &$4d_{5/2}$&$2p_{3/2}$&3.97E-8 & 2.09 &2.69E-21&
4.42E-39&0.258&1.61E18\\
$^{174}$Lu&59.1 &M3&1.23E7 &$4d_{5/2}$&$1s_{1/2}$&9.73E-8 &-4.04 &6.89E-21&
7.63E-39&0.314&3.72E14\\
$^{188}$Re&2.63 &M3&1.12E3 &$4f_{5/2}$&$3p_{1/2}$&4.40E-12&-0.009&6.25E-23&
1.44E-39&0.069&4.76E18\\
\hline
\end{tabular}

\setlength{\oddsidemargin}{-2.5cm}
\newpage
Table II. Two-step deexcitation of isomeric nuclei induced by
XENDT.\\

\begin{tabular}{|c|c|c|c|c|c|c|c|c|c|c|c|c|c|c|c|c|}
\hline
nuclide&$E_{il}$,&E$L$/&$t_i,$ s&$E_{hi},$&E$L$/&$E_{hl}$,&EL/&$B$&
$nl_J$&$n^\prime l^\prime_{J^\prime}$
&$V_{12}$,&$\Delta$, &$P_{tot}$&$\sigma$, &$E_{el}$, &$\Phi_{II},$\\
       &keV      &M$L$ &        &keV     &M$L$  &keV      &ML & &
&&eV       &keV&&cm$^2$&keV&W cm$^{-2}$\\
&&$il$&&&$hi$&&$hl$&&&&&&&&&\\
\hline
$^{58} $Co& 24.9&M3&3.29E4 &28.3&M1&52.9 &E2&2.90E-5&$1s_{1/2}$&$2s_{1/2}$&
6.52E-2 &-21.5&9.16E-11&2.18E-32&12.3&6.57E16\\
$^{99} $Tc&142.6&M4&2.16E4 &38.4&M2&181.1&E2&1      &$1s_{1/2}$&$2p_{3/2}$&
1.50E-4 &-20.1&2.88E-16&9.24E-39&33.7&1.87E19\\
$^{111}$Cd&150.8&E3&2.91E3 &20.5&M2&171.3&M1&1      &$1s_{1/2}$&$2p_{3/2}$&
2.32E-4 &2.69 &3.75E-14&7.45E-37&42.7&2.18E18\\
$^{133}$Ba&275.9&M4&1.40E5 &2.93&E3&278.8&M1&1      &$2p_{3/2}$&$3d_{3/2}$&
6.78E-8 &1.52&9.97E-21&1.03E-41&8.39&6.48E20\\
$^{162}$Ho& 10.0&E3&4.02E3 &65.7&M2& 75.6&M1&1      &$1s_{1/2}$&$2p_{3/2}$&
8.66E-4 &-18.1&1.03E-14&4.74E-38&89.0&5.18E19\\
$^{174}$Lu& 59.1&M3&1.23E7 &29.5&E2& 88.5&M1&1      &$2p_{3/2}$&$3p_{3/2}$&
1.02E-3 &-22.2&2.78E-14&9.23E-36&14.8&1.45E13\\
$^{186}$Re& 50  &E5&6.31E12&37  &M2& 86.6&M3&5.34E-6&$1s_{1/2}$&$2p_{3/2}$&
1.34E-3 &24.1 &1.37E-14&3.79E-38&114.7&9.98E15\\
$^{188}$Re& 15.9&M3&1.12E3 &10.7&E2& 26.7&M1&1      &$2s_{1/2}$&$3d_{5/2}$&
-9.79E-5&-0.026&1.07E-10&9.67E-33&20.0&2.06E14\\
$^{191}$Os& 74.4&M3&4.72E4 &57.6&M1&131.9&E2&2.26E-4&$1s_{1/2}$&$2s_{1/2}$&
8.54E-1 &3.34 &1.55E-7 &4.02E-31&118.2&3.06E15\\
$^{202}$Pb&129.5&E4&1.27E4 &38.6&E2&168.1&E2&0.829  &$2p_{3/2}$&$3p_{3/2}$&
1.82E-3 &-28.6 &4.31E-14&7.20E-36&20.8&3.05E16\\
\hline
\end{tabular}

\newpage
\setlength{\oddsidemargin}{-1cm}
Table III. Emission of M\"{o}ssbauer radiation induced by XENDT.\\

\begin{tabular}{|c|c|c|c|c|c|c|c|c|c|c|c|c|c|c|}
\hline
nuclide&nat. ab./&$E_\gamma$,&E$L$/&half-life,&$nl_J$&$n^\prime
l^\prime_{J^\prime}$&$V_{12}$, &$\Delta$, &$P_{tot}$&$\sigma/
(1+\alpha)$, 
&$E_{el}$, &$N_\gamma/I,$\\
&half-life, s&keV&M$L$&s&&&eV&keV&&cm$^2$&keV&Bq mA$^{-1}$\\
\hline
$^{119}$Sn&8.6  \%&23.9 &M1 &1.80E-8&$1s_{1/2}$&$2s_{1/2}$&0.316 &0.86  &
4.30E-7&7.16E-30&46.7&1.43E4\\
$^{129}$I&4.95E14 &27.8 &M1 &1.68E-8&$1s_{1/2}$&$2s_{1/2}$&0.364 &0.17  &
1.16E-5&1.49E-28&53.1 &5.09E5\\
$^{152}$Eu&4.27E8 &89.8 &E1 &3.84E-7&$1s_{1/2}$&$2p_{1/2}$&1.85  &-48.9 &
3.38E-9&2.04E-32&77.6&5.61E2\\
$^{154}$Eu&2.71E8 &68.2 &E1 &2.2E-6 &$1s_{1/2}$&$2p_{1/2}$&1.86  &-27.3 &
1.10E-8&6.62E-32&77.6&1.42E3\\
$^{153}$Gd&2.09E7 &41.5 &M1 &4.08E-9&$1s_{1/2}$&$2s_{1/2}$&0.571 & 0.30 &
8.64E-6&4.85E-29&80.4&2.51E5\\
$^{157}$Gd&15.7 \%&63.9 &E1 &4.6E-7 &$1s_{1/2}$&$2p_{1/2}$&1.94  &-21.6 &
1.89E-8&1.06E-31&80.4&1.49E3\\
$^{161}$Dy&18.9 \%&25.6 &E1 &2.91E-8&$1s_{1/2}$&$2p_{1/2}$&2.08  &19.6  &
2.64E-8&1.29E-31&86.1&1.13E3\\
$^{161}$Dy&18.9 \%&43.8 &M1 &8.3E-10&$1s_{1/2}$&$2s_{1/2}$&0.614 &0.92  &
1.08E-6&5.28E-30&86.1&2.86E4\\
$^{179}$Ta&5.65E7 &30.7 &E1 &1.42E-6&$1s_{1/2}$&$2p_{1/2}$&2.67  &25.6  &
2.47E-8&7.71E-32&107.8&6.57E2\\
$^{181}$Ta&99.9 \%&6.24 &E1 &6.05E-6&$2p_{3/2}$&$3s_{1/2}$&0.0968&0.94  &
1.09E-7&3.17E-29&15.8&5.17E2\\
$^{189}$Os&16.1 \%&69.6 &M1 &1.62E-9&$1s_{1/2}$&$3s_{1/2}$&0.413 &1.28  &
2.93E-7&7.63E-31&118.2&3.45E3\\
$^{193}$Ir&62.7 \%&73.0 &M1 &6.09E-9&$1s_{1/2}$&$3s_{1/2}$&0.426 &-0.10 &
4.93E-5&1.21E-28&121.8&4.23E5\\
$^{195}$Au&1.61E7 &61.5 &M1 &3.0E-9 &$1s_{1/2}$&$2s_{1/2}$&0.934 &4.91  &
8.79E-8&1.91E-31&129.2&8.41E2\\
$^{197}$Au&100  \%&77.3 &M1 &1.91E-9&$1s_{1/2}$&$3s_{1/2}$&0.454 &-0.054&
1.97E-4&4.29E-28&129.2&3.27E6\\
$^{231}$Pa&1.03E12&84.2 &E1 &4.51E-8&$1s_{1/2}$&$2p_{1/2}$&4.62  &8.07  &
7.22E-7&8.07E-31&180.2&4.97E4\\
$^{237}$Np&6.77E13&102.9&E1 &8.0E-10&$1s_{1/2}$&$2p_{3/2}$&4.88  &-1.91 &
2.83E-5&2.85E-29&189.9&1.57E6\\
\hline
\end{tabular}
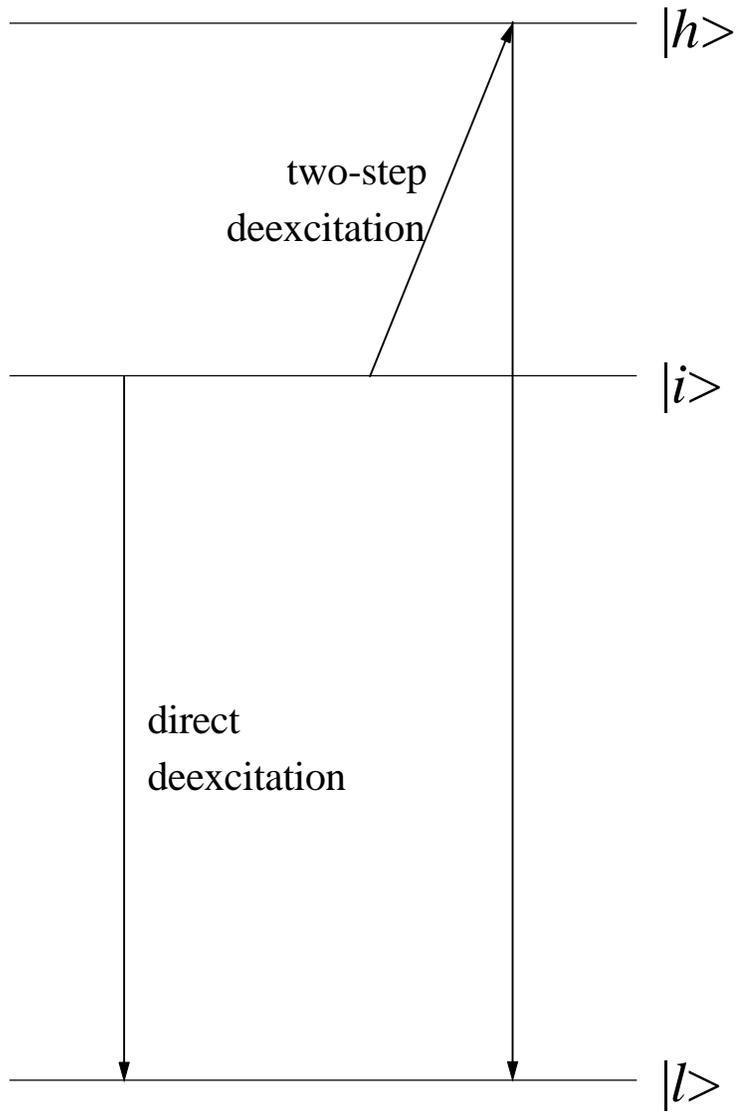
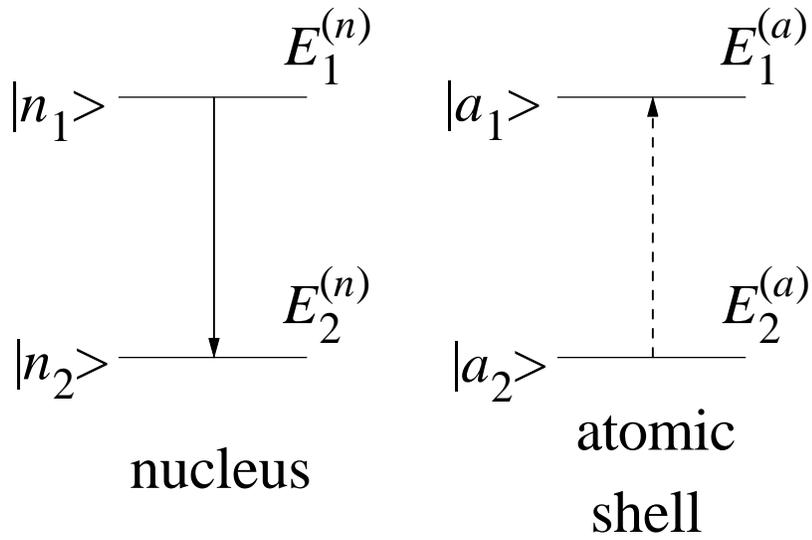
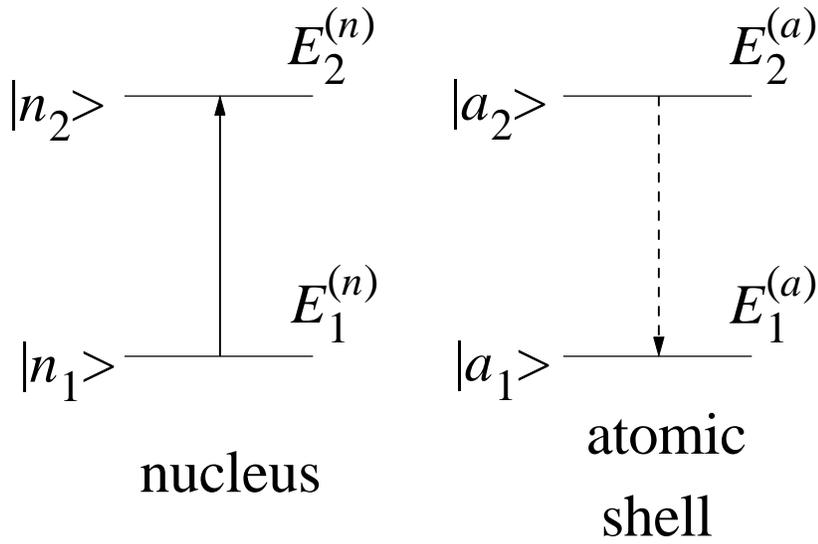
\end{document}